\documentstyle[11pt,aaspp4,epsf,rotate]{article}

\def\psim{\lower.5ex\hbox{$\; \buildrel \propto \over\sim \;$}}
\def\gtrsim{\lower.5ex\hbox{$\; \buildrel > \over\sim \;$}}
\def\lesssim{\lower.5ex\hbox{$\; \buildrel < \over\sim \;$}}
 
\def\g2{\gamma_2} 
\def\tT{\tau_T}
 
\def\e{{\epsilon}}
\def\ag{\alpha_{\gamma}}
\begin{document}

\title{Flash-Heating of Circumstellar Clouds by Gamma Ray Bursts}

\author{Charles D.  Dermer\altaffilmark{1} \& Markus
B\"ottcher\altaffilmark{2,3}}

\altaffiltext{1}{E.  O.  Hulburt Center for Space Research, Code 7653,
Naval Research Laboratory, Washington, DC 20375-5352}
\altaffiltext{2}{Department of Space Physics and Astronomy, Rice
University, Houston, TX 77005-1892} \altaffiltext{3}{Chandra Fellow}

\begin{abstract}

The blast-wave model for gamma-ray bursts (GRBs) has been called into
question by observations of spectra from GRBs that are harder than can
be produced through optically thin synchrotron emission.  If GRBs originate 
from the collapse of massive stars, then circumstellar clouds near burst
sources will be illuminated by intense $\gamma$ radiation, and the
electrons in these clouds will be rapidly scattered to energies as
large as several hundred keV.  Low-energy photons that subsequently
pass through the hot plasma will be scattered to higher energies, thus
hardening the intrisic spectrum.  This effect resolves the
``line-of-death'' objection to the synchrotron shock model. Illuminated
clouds near GRBs will form relativistic plasmas containing large
numbers of electron-positron pairs that can be detected within $\sim$
1-2 days of the explosion before expanding and dissipating.  Localized
regions of pair annihilation radiation in the Galaxy would reveal past
GRB explosions.  \end{abstract}


\section{Introduction}

The identification of flaring and fading X-ray, optical and radio
counterparts to gamma-ray burst (GRB) sources (e.g., Costa et al.\
\markcite{cea97}1997; van Paradijs et al.\ \markcite{vpea97}1997;
Djorgovski et al.\ \markcite{dea97}1997; Frail et al.\
\markcite{fea97}1997), and the large energy releases implied by
redshift measurements, find a consistent explanation in an expanding
relativistic blast-wave model (Paczy\'nski \& Rhoads
\markcite{pr93}1993; M\'esz\'aros \& Rees \markcite{mr97}1997).  As a
result of Beppo-SAX and optical follow-on observations, the redshifts
of about one dozen GRBs with durations greater than $\sim 1$ s have
been measured.  The distribution of redshifts is broad and centered
near $z\sim 1$, corresponding to the cosmological epoch of active star
formation (Hogg \& Fruchter \markcite{hf99}1999).  GRBs are extremely
luminous and energetic at hard X-ray and $\gamma$-ray energies.  The
degree of GRB collimation is unknown, but peak directional
$\gamma$-ray luminosities and energy releases as large as $\partial
L/\partial \Omega\simeq 3\times 10^{51}$ ergs (s-sr)$^{-1}$ and $\partial
E/\partial \Omega \simeq 3\times 10^{53}$ ergs sr$^{-1}$, respectively,
have been measured (Kulkarni et al.\ \markcite{kea99}1999).  Less
powerful GRBs and less luminous episodes during the GRB produce
smaller $\gamma$-ray powers, but the apparent isotropic $\gamma$-ray
luminosities from typical GRBs could regularly reach values exceeding
$10^{50}L_{50}$ ergs s$^{-1}$ with $L_{50} \sim 1$, with some GRBs
reaching $L_{50} > 10^2$.  Because the energy radiated in $\gamma$
rays is less than the total energy released by a GRB, the apparent
isotropic energy release of GRB sources could often reach values of
$10^{54}E_{54}$ ergs, with $E_{54} \sim$1.

Cosmological gamma-ray burst and afterglow observations are best
explained through the fireball/blast-wave model, where the deposition
of large quantities of energy into a small region yields a fireball
that expands until it reaches a relativistic speed determined by the
amount of baryons mixed into the fireball (see, e.g., Piran
\markcite{tp99}1999 for a review).  Nonthermal synchrotron radiation
from energetic electrons in the relativistic blast wave is thought to
account for the origin of the prompt $\gamma$-ray emission and
afterglow radiation.  This paradigm has been called into question,
however, by observations of very hard X-ray emission during the prompt
$\gamma$-ray luminous phase of a significant number of GRBs (Crider et al.\
\markcite{crider97}1997; Preece et al.\ \markcite{pea98}1998).  Photon
fluxes $\phi(\e ) \propto \e^{-\alpha_X}$ with $\alpha_X \sim$ 0,
where $\e = h\nu/m_ec^2$ is the dimensionless photon energy, have been
observed in 5-10\% of GRBs that are bright enough to permit spectral
analysis.  This strongly contradicts the optically-thin synchrotron 
shock model, which predicts that only radiation spectra with 
$\alpha_X \geq 2/3$ can emerge from the blast wave. In view 
of the severity of this challenge to the model, these observations 
have been termed the ``line-of-death" to the synchrotron shock model.  
Possible explanations for this phenomeonon involve photoelectric 
absorption by optically thick cold matter (Liang \& Kargatis
\markcite{lk94}1994; Brainerd \markcite{jb94}1994; B\"ottcher 
et al.\ \markcite{bea99}1999), synchrotron self-absorption 
(Crider \& Liang \markcite{cl99}1999, Granot, Piran, \& Sari
\markcite{granot00}2000, Lloyd \& Petrosian \markcite{lp99}1999), 
Compton scattering (Liang \markcite{liang97}1997; Liang et al.\ 
\markcite{lcbs99}1999), or the existence of a
pair-photosphere (M\'esz\'aros \& Rees \markcite{mr00}2000) within the
blast wave.  Except for the last model cited, these explanations
are inconsistent with the standard synchrotron shock model.  Here we offer a
solution to this problem that is consistent with the standard model
and recent observations pointing to a massive star origin of GRBs.

\section{Massive Star Origin of GRBs}

Considerable evidence linking the sources of GRBs with star-forming
regions has recently been obtained (e.g., Lamb \markcite{lamb99}1999).
For example, the associated host galaxies have blue colors, consistent
with galaxy types that are undergoing active star formation.  GRB
counterparts are found within the optical radii and central regions of
the host galaxies (e.g.  Bloom et al.\ \markcite{bea99a}1999a), rather
than far outside the galaxies' disks, as might be expected in a
scenario of merging neutron stars and black holes (Narayan, Paczy\'nski,
\& Piran \markcite{npp92}1992).  Lack of optical counterparts in some
GRBs could be due to extreme reddening from large quantities of gas
and dust in the host galaxy.  This, together with the appearance of
supernova-like emissions in the late time optical decay curves of a
few GRBs (e.g., Bloom et al.\ \markcite{bea99b}1999b) and weak X-ray
evidence for Fe K$_\alpha$-line signatures (Piro et al.\
\markcite{pea99}1999), supports a massive star hypernova/collapsar
(Woosley \markcite{sw93}1993; Paczy\'nski \markcite{bp98}1998)
origin for the long duration gamma-ray bursters.

The observations thus favor a model for GRBs involving the collapse of
the core of a $\gtrsim 30 M_\odot$ star to a black hole, with the
collapse events producing fireballs and relativistic outflows with
large directed energy releases.  Earlier treatments of the blast-wave
model considered systems where the density of the surrounding medium
is either uniform or monotonically decreasing as a result of a
circumstellar medium formed by a hot stellar wind (M\'esz\'aros, Rees,
\& Wijers \markcite{mrw98}1998).  Until recently (Chevalier \&
Li \markcite{cl99}1999; Li \& Chevalier \markcite{lc99}1999),
less attention has been paid to the actual 
environment found in the vicinity of massive stars.  For this
we consider $\eta$ Carinae (Davidson \& Humphreys
\markcite{dh97}1997), the best-studied massive star that might
correspond to a GRB progenitor.  It is an evolved star with
present-day mass $\geq 90 M_\odot$, distance of $2300\pm 200$ pc, and
lifetime of about 3 million years.  It anisotropically ejects mass at
a current rate of $\lesssim 0.003 M_\odot$ yr$^{-1}$ to form its
unusual ``homonculus nebula."  Several Solar masses of material
surround $\eta$ Carinae.  In the immediate vicinity of the central
star, dense clouds of slow-moving gas with radii $r \sim 10^{15}$ cm
and densities between $10^7$ and $10^{10}$ cm$^{-3}$ were discovered
with speckle techniques (Hofman \& Weigelt \markcite{hw88}1988) and
confirmed with high-resolution HST observations (Davidson et al.\
\markcite{dea95}1995).  This material, moving with speeds of $\sim 50$
km s$^{-1}$, is apparently ejected nonuniformly from the equatorial
zone, but may remain trapped by the gravitational field of the star.
Inferences (Davidson \& Humphreys \markcite{dh97}1997) from [FeII]
observations suggest that $\gtrsim 0.02 M_\odot$ of gas are contained
within $\sim 2\times 10^{16}$ cm, implying a volume-averaged gas
density $\gtrsim 7\times 10^5$ cm$^{-3}$.  Model results (B\"ottcher
\markcite{mb99}1999) imply that a dense ($n \sim 10^{12}$ cm$^{-3}$)
torus of gas at mean distance $d\sim 2\times 10^{15}$ cm and with a
10-fold enhancement of Fe relative to Solar abundance is required to
explain the Fe K$_\alpha$ emission weakly detected from GRB 970508
with Beppo-SAX (Piro et al.\ \markcite{pea99}1999).

Guided by the optical observations of $\eta$ Carinae, we assume that
the volume-averaged density of gas at $d \lesssim 10^{16}$ cm of a GRB
source is $\langle n \rangle = 10^6 n_6$ cm$^{-3}$.  Dense clouds of
radius $10^{15}r_{15}$ cm and radial Thomson depths $\tT = n_c\sigma_T
r$ are assumed to be embedded within this region, so that the mean
density of particles in a cloud is $n_c = 1.5\times 10^9 \tT /r_{15}$
cm$^{-3}$.  Thus $\tT \sim 1$ clouds located very close to a GRB
source are consistent with the observations of dense blobs near $\eta$
Carinae.  The deceleration length scale of a blast wave with initial
Lorentz factor $\Gamma_0 = 100\Gamma_2$ in a uniform medium is $x_d =
(3E_0/4\pi \Gamma_0^2 \langle n \rangle m_pc^2)^{1/3} = 2.5\times
10^{15}(E_{54}/\Gamma_2^2 n_6)^{1/3}$ cm; hence the blast wave would
emit a significant fraction of its energy before reaching distances of
$\sim 10^{16}$ cm.  The deceleration time, which corresponds to the
duration of the prompt $\gamma$-ray luminous phase of a GRB in the
external shock model (Rees \& M\'esz\'aros \markcite{rm92}1992), is
$t_d = (1+z) x_d/(c \Gamma_0^2) \cong 8(1+z) (E_{54}/\Gamma_2^8
n_6)^{1/3}$ s.  These parameters are not unique, and we expect that
GRBs display a wide range of energies, Lorentz factors and surrounding
mean densities that could accommodate the diverse range of GRB
observations.

\section{Blast-Wave/Cloud Interaction}

A wave of photons impinging on a cloud located $10^{16} d_{16}$ cm
from the explosion center will photoionize and Compton-scatter the
ambient electrons to energies characteristic of the incident $\gamma$
rays (Madau \& Thompson \markcite{mt99}1999).  The $\gamma$-ray photon
front has a width of $\sim$10-100 lt-s, corresponding to the duration
of the GRB, whereas the plasma cloud has a width of $\sim 3\times 10^4
r_{15}$ lt-s, so that the radiation effects must be treated locally.
The radiation force driving the electrons outward is balanced by
strong electrostatic forces from the more massive protons and ions
that anchor the system until the net impulse is sufficient to drive
the entire plasma cloud outward.  The Compton back-scattered photons
provide targets for successive waves of incident GRB photons through
$\gamma\gamma$ pair-production interactions (Thompson \& Madau
\markcite{tm99}1999).  Higher-energy photons are preferentially
attenuated, forming an additional injection source of $\gtrsim 1$ MeV
electron-positron pairs.  The nonthermal electrons and pairs will
Compton scatter successive waves of photons, thereby modifying the
incident spectrum.  The pairs, no longer bound by electrostatic
attraction with the ions, will be driven outward by both radiation
forces and restoring electrostatic fields to form a mildly
relativistic pair wind passing through the more slowly moving normal
plasma.  Shortly after the $\gamma$-ray photon front has passed, the
decelerating blast wave from the GRB will plow into the cloud,
shock-heating the relativistic plasma.

Nonthermal synchrotron photons with energy $\e$ impinge on the atoms
in the cloud with a flux which can be parametrized as
\begin{equation} 
\Phi(\e) = (4\pi d^2)^{-1}\; {L \over m_ec^2 \e_0^2 \zeta_1}
\;\big[{1\over (\e/\e_0)^{2/3}+(\e/\e_0)^{\alpha_\gamma}}\bigr]\; 
\label{Phi}
\end{equation}
(Dermer, Chiang, \& B\"ottcher \markcite{dcb99}1999), where $\e_0 \sim
1$ is the photon energy of the peak of the $\nu F_\nu$ spectrum,
$\alpha_\gamma \sim 2$-3 is the photon spectral index at energies
$\e\gg \e_0$, and $\zeta_1 \cong [3/4 + (\ag-2)^{-1}]$.  Hydrogen, the most
abundant species in the cloud, will be ionized on a time scale of
$5\times 10^{-5} (1+z) d_{16}^2 \zeta_1 \e_0^{4/3}/L_{50}$~s.  Fe
features might persist briefly during the early periods of very weak
GRBs on a time scale of $4\times 10^{-3} (1+z) d_{16}^2 \zeta_1
\e_0^{4/3}/L_{50}$~s, and would be identified by a rapidly evolving Fe
absorption feature at $9.1/(1+z)$ keV.  After the H and Fe are
ionized, the coupling between the GRB photons and gas is dominated by
Compton scattering interactions.  Pair production through
photon-particle processes are negligible by comparison with Compton
interactions except for photons with $\e \gtrsim 200$.  A lower limit
to the time scale for an electron to be scattered by a photon is
$t_T(s) \approx 15 (1+z) d_{16}^2 \zeta_1\e_0/\zeta_2 L_{50}$, where
$\zeta_2 = [3 + (\alpha_\gamma -1)^{-1}]$, assuming that all Compton
scattering events occur in the Thomson limit.  The Klein-Nishina 
decline in the Compton cross section will increase this
estimate by a factor of $\sim 1$-3, depending on the incident
spectrum.  Most of the electrons in the cloud will therefore be
scattered to high energies during a very luminous ($L_{50}\gg 1$) GRB,
or when the cloud is located at $d_{16} \ll 1$.

The average energy transferred to an electron at rest when
Compton-scattered by a photon with energy $\e$ is $\Delta\e \cong
\e^2/(1+1.5\e )$ (this expression is accurate to better than 18\% for
$\e < 10^3$).  Defining $\eta = \gamma -1$ as the dimensionless
electron kinetic energy, we can easily estimate the production rate
$f(\eta )$ of electrons scattered to energy $\eta$ in the
nonrelativistic ($\eta \ll 1$) and extreme relativistic ($\eta \gg 1$)
limits, noting that $f(\eta)d\eta \propto \Phi(\e)\sigma_C(\e)d\e$ and
letting $\eta\cong \Delta\e$.  In the former limit, the Compton cross
section $\sigma_C(\e)\rightarrow \sigma_T$ and $\Phi(\e) \propto
\e^{-2/3}$, so that $f(\eta)\propto \eta^{-5/6}$ when $\eta\ll$ min(1,
$\e_0^2$).  In the high energy limit, $\sigma_C(\e)\propto
\ln(3.3\e)/\e$ and $\Phi(\e) \propto \e^{-\alpha_\gamma}$, so that
$f(\eta )\propto \eta^{-(\alpha_\gamma+1)} \ln(2.2\eta)$ when
$\eta\gg$ max$(1, \e_0^2$).  Thus electrons are Compton-scattered on
the time scale derived above to form a hard spectrum that turns over
at kinetic energies of $\gtrsim 500\times$min($1,\e_0^2$) keV.  For a
GRB with $\e_0 \sim 1$, most of the kinetic energy is therefore
carried by nonthermal electrons with energies of $\sim 500$ keV.
Successive waves of photons that pass through this plasma will
continue to Compton-scatter the nonthermal electrons.  Only the lowest
energy photons, however, will be strongly affected by the radiative
transfer because both the Compton scattering cross section and energy
change per scattering is largest for the lowest energy photons.

Following the initial wave of photons, successive photon fronts also
encounter the back-scattered radiation (Madau \& Thompson
\markcite{mt99}1999; Thompson \& Madau \markcite{tm99}1999).  The
kinematics of the Compton process dictate that the energy $\e_s$ of a
photon back-scattered through $180^\circ$ by an electron at rest is
$\e_s = \e/(1+ 2\e)$; thus $\e_s$ cannot exceed 1/2 the electron
rest-mass energy.  Head-on collisions of the back-scattered photons by
primary GRB photons with $\e_1 > 2/\e_s = 2+2\sqrt{3}$ can thus produce
nonthermal e$^+$-$e^-$ pairs.  The cross section for $\gamma\gamma$
pair production peaks near threshold with a value of $\sim \sigma_T/3$.
The $\gamma\gamma$ pair-production optical depth $\tau_{\gamma\gamma}
(\e_1)$ of a photon that trails the onset of the GRB by $\Delta t$
seconds can be estimated by noting that the photon traverses a
distance $\sim r$ through a backscattered radiation field of spectral
density $n_s(\e_s)\approx n_e \sigma_T \cdot \Delta t \cdot \Phi [\e_s
/(1-2\e_s )]$ -- a more accurate calculation would replace the term
$\Delta t \cdot \Phi [\e_s/(1-2\e_s)]$ by an integral over the
time-varying flux.  Approximating $\tau_{\gamma\gamma} (\e_1)\approx r
(\sigma_T/3) \Delta\e_s n_s(2/\e_1)$, where $\Delta \e_s \simeq 2/\e_1$ is
the bandwidth that is effective for producing pairs, we obtain
\begin{equation} 
\tau_{\gamma\gamma} (\e_1) \approx 0.02 \; {\tau_T \Delta t [s] \, 
L_{50} k(\e_1) \over d_{16}^2 \e_1\e_0^2\zeta_1}\;\big[{1\over 
(\e^\prime/\e_0)^{2/3} + (\e^\prime/\e_0)^{\alpha_\gamma}}\bigr]\; , 
\label{tau_gg}
\end{equation}
where $\e^\prime = 2/(\e_1-4)$.  The coefficient results from a more
detailed derivation, and the term $k(\e_1) = 1-4\e_1^{-1} + \e_1
/(\e_1 -4)$ is a Klein-Nishina correction.  Eq.\ (\ref{tau_gg}) shows 
that photons with energies above several MeV will be severely attenuated 
in Thomson thick clouds if $L_{50}\gg 1$ or $d_{16}\ll 1$.  Photons with
MeV energies are most severely attenuated, and $\tau_{\gamma\gamma}
(\e_1)\propto \e_1^{-1/3}$ at energies $\e_1 \gg \max (1, \e_0)$.
The $\gamma\gamma$ pair injection process provides another source of
nonthermal leptons with $\eta \sim 1$.  The pairs will not, however,
be electrostatically bound but will be accelerated by the photon
pressure and electrostatic field.  

Fig.\ 1 shows Monte Carlo simulations of radiation spectra 
described by eq.\ (\ref{Phi}) that pass through a hot electron 
scattering medium.  For simplicity, we approximate the 
hard nonthermal electron spectrum by a thermal
distribution with temperatures of 100 and 300~keV.  These
calculations show that the lowest energy photons of the primary
synchrotron spectrum are most strongly scattered, and that the 
``line-of-death" problem of the synchrotron shock model of 
GRBs (Preece et al.\ \markcite{pea98}1998) can be solved by 
radiation transfer effects through a hot scattering cloud 
with $\tT \gtrsim$ 1-2.  

According to this interpretation, GRBs 
displaying very hard spectra could display one break from the 
intrinsic synchrotron shock spectrum and a second break
from the scattering process. In the examples shown in Fig. 1,
these two breaks are so close to each other that they appear
as one smooth turnover. Two breaks are observed from GRB
970111 (Crider \& Liang \markcite{cl99}1999), a GRB that strongly
violates the ``line of death." A prediction of this model is
that GRBs showing such flat X-ray spectra should also display softer
MeV spectra than typical GRBs due to $\gamma\gamma$ attenuation
processes in the hot scattering cloud.

\section{Observational Signatures of the Flash-Heated Cloud}

The Compton-scattered electrons transfer momentum to the $N_p =
4\times 10^{54}r_{15}^3n_9$ protons of the cloud through their
electrostatic coupling.  If the radiation efficiency is $\xi_r$, then
the Compton impulse gives each proton in the cloud $\approx \xi_r
[1-\exp(-\tT )]E \pi r^2/(4\pi d^2N_p) \approx 40 [1-\exp(-\tT )]
(\xi_r/0.1) E_{54}/(d_{16}^2 r_{15}n_9)$ MeV of directed energy.
Pairs, by contrast, will be accelerated to mildly relativistic speeds
until Compton drag or streaming instabilities limit further
acceleration.  In the simplification that the medium interior to the
cloud is uniform, and neglecting pair-loading of the swept-up material
(Thompson \& Madau \markcite{tm99}1999), the decelerating blast wave
follows the dynamical equation $\Gamma(x) = \Gamma_0(x/x_d)^{-g}$,
where $g = 3/2$ and 3 for adiabatic and radiative blast waves,
respectively.  Using the standard parameters adopted here, the blast
wave slows to between $\xi= 0.01 - 0.1$ of its initial Lorentz factor
before reaching a cloud at $d = 10^{16}$ cm.  Even considering the
radiative acceleration of the cloud, the blast wave reaches the cloud
at time $t_{bw} = t_d(d/x_d)^{(2g+1)}/(2g+1) \lesssim t_{\rm dyn}$,
where the dynamical time scale of the cloud is $t_{\rm dyn} = r/c =
3\times 10^4 r_{15}$ s.  Because the cloud is so dense, a large
fraction of the residual energy of the blast wave is deposited into
the $N_p$ particles of the cloud.  Thus each proton in the cloud
receives an additional $m_p\beta_p^2 c^2 \approx \xi E \pi r^2/(4\pi
d^2N_p) \approx 40 (\xi/0.1) E_{54}/(d_{16}^2 r_{15}n_9$) MeV of
kinetic energy, divided roughly equally into directed outflow and
random thermal energy.

If the circumstellar medium at $d_{16}\gg 1$ is much more dilute than
the interior region, as suggested by observations of $\eta$ Carinae,
then we can neglect further interactions of the cloud/blast wave
system with their surroundings.  The observational signatures and fate
of the cloud at late times can be outlined by comparing time scales.
The cloud expands on a time scale $t_{\rm ex} = t_{\rm dyn}/\beta =
1.5\times 10^5 r_{15}/(\beta/0.2)$ s.  The basic time scale
governing radiative processes in the cloud is the Thomson time $t_{\rm
T} = (n_e\sigma_T c)^{-1} = 5\times 10^4/n_9$~s $= t_{\rm dyn} /\tau_{\rm
T}$.  The electrons thermalize on a time scale $t_{\rm T}/\ln\Lambda
\ll t_{\rm dyn}$, where the Coulomb logarithm $\ln\Lambda \approx 20$.
The protons transfer their energy to the electrons on the time
scale $t_{ep}\cong \theta^{3/2}(m_p/m_e)t_{\rm T}/\ln\Lambda$, 
where $\theta = kT/m_ec^2$ is an effective dimensionless
electron temperature, and we assume collective plasma processes for
energy exchange to be negligible.

The flash-heated cloud can evolve in two limiting regimes.  When the
external soft-photon energy density is small, the system emits by a
hard bremsstrahlung spectrum with $\theta \sim$ 1 and luminosity
$L_{ff}\cong N_e\alpha_f m_ec^2\theta^{1/2}/t_{\rm T} \approx 5\times
10^{41} r_{15}^3 n_9^2\theta^{1/2}$ ergs s$^{-1}$.  In the more likely
case when abundant soft photons are present, for example, from the
Compton echo (Madau, Blandford, \& Rees \markcite{mbr99}1999), then
Compton cooling will balance ion heating to produce a luminous
Comptonized soft-photon spectrum with effective temperature $\theta
\sim 0.1$ and luminosity $L_{\rm C}\cong \xi E \pi r^2/(4\pi d^2
t_{ep}) \approx 2\times 10^{45} (\xi/0.1) E_{54} r_{15}^2 n_9
d_{16}^{-2}(\theta/0.1)^{-3/2}$ ergs s$^{-1}$.  In either case, the
spectra persist until the plasma expands and adiabatically cools, that
is, for a period $\sim t_{ex} \sim$ day.  The hot bremsstrahlung
plasma will be too dim to be detectable with current instrumentation,
but a 50-100 keV Comptonized plasma at redshift $z \sim 1$ and luminosity 
distance of $10^{28}D_{28}$ cm would have
a flux of $\sim 1.6\times 10^{-12} (\xi/0.1) E_{54} r_{15}^2 n_9
d_{16}^{-2}(\theta/0.1)^{-3/2} D_{28}^{-2}$ ergs cm$^{-2}$ s$^{-1}$.
The hot plasma formed by a nearby GRB at $z \sim 0.1$ would be
 easily detectable with the
INTEGRAL and Swift missions at hard X-ray and soft $\gamma$-ray
energies.  In either case, e$^+$-e$^-$
pairs would be formed with moderate efficiency, 
and the cooling, expanding plasma
would produce a broad pair annihilation feature (Guilbert \& Stepney
\markcite{gs85}1985).  The residual pairs formed in the relativistic
plasma and the pair wind would diffuse into the dilute interstellar
medium with density $n_{\rm ISM}$ to annihilate on a time scale
$(n_{\rm ISM} \sigma_{\rm T} c)^{-1} \sim 2 \times 10^6/n_{\rm ISM}$
yr.  If the energy intercepted by a single cloud is converted to pairs
with a conservative 1\% pair yield, past GRBs in the Milky Way would
be revealed by localized regions of annihilation radiation with flux
$\sim 2\times 10^{-5} E_{54} n_{\rm ISM} ~(d/10 {\rm kpc})^{-2}$ 0.511
MeV ph cm$^{-2}$ s$^{-1}$.  The high-latitude annihilation feature
discovered with OSSE on the Compton Gamma Ray Observatory (Purcell et
al.\ \markcite{pea97}1997), or other localized hot spots of
annihilation radiation that will be mapped in detail with INTEGRAL,
could reveal sites of past GRB explosions. 

\acknowledgments{CD thanks B.  Paczy\'nski for stressing the importance
of massive star observations in developing blast-wave models of GRBs.
The work of CD is supported by the Office of Naval Research and NASA
Astrophysical Theory Program (DPR S-13756G). The work of MB is
supported by NASA through Chandra Postdoctoral Fellowship grant
PF~9-10007, awarded by the Chandra X-ray Center, which is operated
by the Smithsonian Astrophysical Observatory for NASA under
contract NAS~8-39073.}

\eject

\setcounter{figure}{0}
\begin{figure} 
\rotate[r]{ \epsfysize=12cm
\epsffile[150 0 550 500]{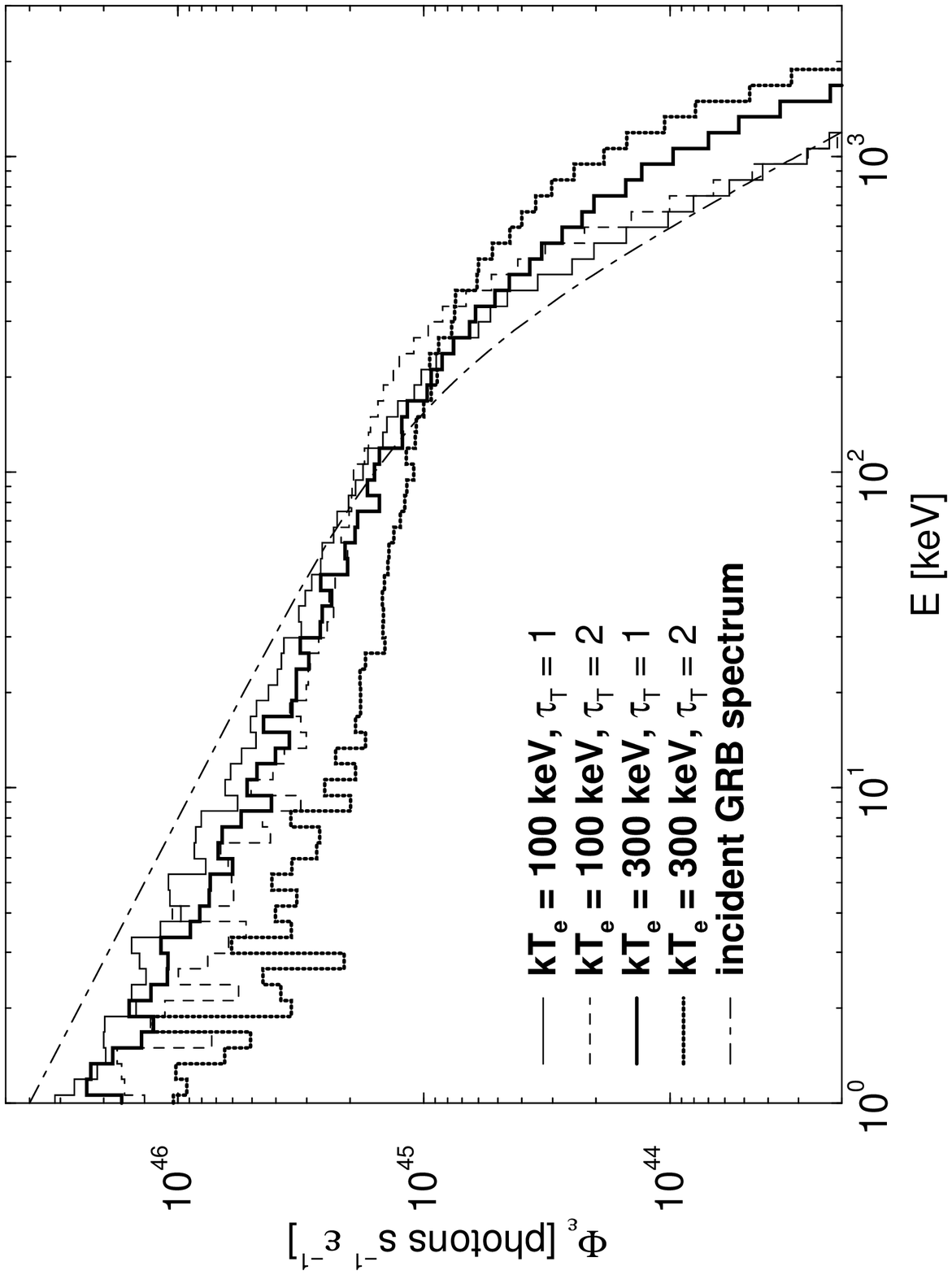} }

\caption[]{Radiation transfer
effects on GRB emission that passes through electrons energized by an
earlier portion of the photon front.  The intrinsic spectrum eq.(1),
with $\alpha_X = 2/3$, $\alpha_\gamma = 2.5$ and $\e_0 = 0.5$, is
shown by the thick dashed curve.  The nonthermal electrons are
approximated by a thermal distribution with temperatures of 100 keV
(thin curves) and 300 keV (thick curves), and Thomson depths $\tT = 1$
(solid curves) and $\tT = 2$ (dotted curves).  Spectral indices
$\alpha_X$ calculated at 50 keV are 0.5 ($T = 100$ keV, $\tT = 1$),
0.44 ($T = 300$ keV, $\tT = 1$), 0.14 ($T = 100$ keV, $\tT = 2$), and
0.05 ($T = 300$ keV, $\tT = 2$).  } 
 
\label{figure1} 
\end{figure}
\end{document}